\pdfoutput=1
\documentclass[aps,prl,twocolumn,superscriptaddress,showpacs]{revtex4}

\usepackage[pdftex]{graphicx}
\usepackage{amsmath,amssymb}

\begin{document}

\title{Unidirectional band gaps in uniformly magnetized two-dimensional magnetophotonic crystals}

\author{Mathias Vanwolleghem}\email[]{mathias.vanwolleghem@u-psud.fr}
\affiliation{Institut d'Electronique Fondamentale, CNRS,
Universit\'e Paris-Sud, Centre Scientifique d'Orsay, F-91405 Orsay, France}
\author{Xavier Checoury}
\affiliation{Institut d'Electronique Fondamentale, CNRS,
Universit\'e Paris-Sud, Centre Scientifique d'Orsay, F-91405 Orsay, France}
\author{Wojciech \'Smigaj}
\affiliation{Institut Fresnel, CNRS, Aix-Marseille Universit\'e, Ecole Centrale Marseille,
Campus de St J\'er\^ome, 13397 Marseille Cedex 20, France}
\author{Boris Gralak}
\affiliation{Institut Fresnel, CNRS, Aix-Marseille Universit\'e, Ecole Centrale Marseille,
Campus de St J\'er\^ome, 13397 Marseille Cedex 20, France}
\author{Liubov Magdenko}
\affiliation{Institut d'Electronique Fondamentale, CNRS,
Universit\'e Paris-Sud, Centre Scientifique d'Orsay, F-91405 Orsay, France}
\author{Kamil Postava}
\affiliation{Department of Physics, Technical University Ostrava, 70833 Ostrava-Poruba, Czech Republic}
\author{B\'eatrice Dagens}
\affiliation{Institut d'Electronique Fondamentale, CNRS,
Universit\'e Paris-Sud, Centre Scientifique d'Orsay, F-91405 Orsay, France}
\author{Pierre Beauvillain}
\affiliation{Institut d'Electronique Fondamentale, CNRS,
Universit\'e Paris-Sud, Centre Scientifique d'Orsay, F-91405 Orsay, France}
\author{Jean-Michel Lourtioz}
\affiliation{Institut d'Electronique Fondamentale, CNRS,
Universit\'e Paris-Sud, Centre Scientifique d'Orsay, F-91405 Orsay, France}

\date{\today}

\begin{abstract}
By exploiting the concepts of magnetic group theory we show how unidirectional behavior can be obtained in two-dimensional magneto-photonic crystals (MOPhC) with uniform magnetization. This group theory approach generalizes all previous investigations of one-way MOPhCs including those based on the use of antiparallel magnetic domains in the elementary crystal cell. Here, the theoretical approach is illustrated for one MOPhC example where unidirectional behavior is obtained by appropriately lowering the geometrical symmetry of the elementary motifs. One-way transmission is numerically demonstrated for a photonic crystal slice.
\end{abstract}


\maketitle

The unique dispersion properties of photonic crystals (PhC) have led over the past decades to the theoretical and experimental demonstration of many unusual and extraordinarily enhanced optical phenomena \cite{Joannopoulosbook, Lourtiozbook}. Photonic crystals made of magneto-optical (MO) materials, where time reversal symmetry is broken, open a new research field on non-reciprocal behavior in optics. Several papers have thus proposed theoretical non-reciprocal optical demonstrators based on such MO PhCs. Indeed, time reversal breaking through the presence of MO materials is greatly enhanced by exploiting the strong dispersion of photonic crystals. In regions of low group velocity, this can lead to "unidirectional freezing" of light \cite{Takeda:PRA08,Figotin:PRB03}: forward light propagates in the crystal with finite velocity whereas backward light is stopped. Recently, Yu et al. have even shown that time reversal breaking can lead to a non-reciprocal shift of bandgap edges in a one-dimensional periodic MO Bragg stack \cite{Yu:APL07}. In this case, there appears a frequency range within which light can freely enter the MO PhC in one sense, while it is totally reflected in the opposite sense. However, Figotin et al. have shown that optical unidirectionality requires both time reversal and space reversal breaking of the structure \cite{Figotin:PRB03}. Till now the latter condition has only been achieved in theory by using periodic structures with antiparallel magnetic domains in the unit cell. Such structures would require a very precise control of magnetic domain walls.

In this Letter we show how, by exploiting the concepts of magnetic group theory, one-way behavior can be obtained in a generic manner in two-dimensional MO PhC's. A general scheme can thus be used to classify all geometries of a magnetophotonic crystal with possible unidirectional effects. In essence unidirectionality requires first and foremost that the symmetry group of the periodic structure be reduced in such a manner that an arbitrary Bloch k-vector of the Brillouin zone (BZ) and its time reversed opposite are not equivalent. For this, time reversal breaking is a necessary but not a sufficient condition. In general, one has to imagine a crystal layout in which none of the present symmetries allows the transformation $\mathbf{k} \rightarrow -\mathbf{k}$. In what follows, we illustrate how group theory concepts can be used to obtain non-reciprocity in a uniformly magnetized PhC by simply reducing the symmetry of the elementary motifs. Theoretical predictions are then numerically validated by demonstrating a pronounced one-way transmission through a finite slice of the crystal. We will finally indicate how properly choosing a 2D MO PhC layout among all possible magnetic groups can generalize such a one-way behavior. 

\begin{figure}[!htb]
\includegraphics[width=\columnwidth]{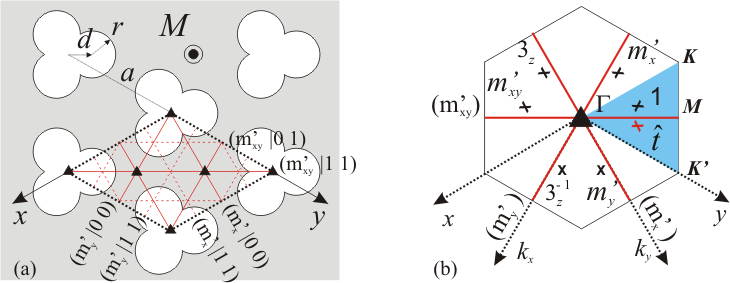}
\caption{(a) Uniformly magnetized hexagonal magnetophotonic crystal with a motif of triangular symmetry. Symmetry elements indicated in the unit cell (dotted black lines) are: rotation centres of order 3 (black triangles), anti-mirror planes (time reversal + mirror operation - continuous red lines) and anti-glide planes (dashed red lines). (b) First Brillouin zone (hexagon) and its irreducible part (IBZ) (shaded triangle $\Gamma-K-K'$). Black crosses form the ``star representation'' of an arbitrary $\mathbf{k}$-vector. Both a Bloch mode and its time reversed counterpart (red cross) are within the IBZ.\label{figlayout}}
\end{figure}
Fig.\ref{figlayout}(a) shows the specific (but not exclusive) layout that we have chosen to illustrate the general concept. A clover-like airhole motif repeated on a hexagonal lattice is etched into a MO background material that is magnetized parallel to the air holes. The possibility of achieving band structure non-reciprocity for this 2D MOPhC with uniform magnetization can be understood from the analysis of the crystal symmetries. Because the axial vector $\mathbf{M}$ is reversed under time reversal, time reversal is not a symmetry operation for Maxwell's equation. As a result, the time-reversed Bloch mode ($\mathbf{k} \rightarrow -\mathbf{k}$) of an arbitrary $\mathbf{k}$-vector in the Brillouin zone (BZ) will \textit{a priori} not be a member of the $\mathbf{k}$-star representation of this solution \cite{Bradley:RevModPhys68} (Fig.\ref{figlayout}(b)). This is a first step towards a non-reciprocal band structure ($\omega_{n,\mathbf{k}}\neq\omega_{n,-\mathbf{k}}$), but other symmetry operations may still be present that transform a given $\mathbf{k}$-vector into its opposite. In order to precisely determine the general $\mathbf{k}$-star of the BZ, one needs to identify the ``magnetic'' space group elements of the MOPhC. The mathematical derivation of this space group is outside the scope of this paper. Using the construction derived by Indebom \cite{Indenbom} one proves that the plane magnetic 2D group $\mathcal{M}$ of the MOPhC of Fig.\ref{figlayout} is given by

\begin{equation}
\mathcal{M} = p3 + \hat{t}(p3m1-p3) = p3m'1\label{groupequation1}
\end{equation}
\label{groupequation}
where $\hat{t}$ is the anti-unitary identity operator (i.e. the time inversion operator, $\hat{t}\mathbf{M}=-\mathbf{M}$). $p3$ and $p3m1$ are the conventional notations for the plane triangular groups \cite{Opechowski} and the prime stands for combination with time reversal ($\hat{t}$). Fig.\ref{figlayout}(a) illustrates the symmetry elements of $p3m'1$. The unitary rotations of order 3 (subgroup $p3$) are indicated by black triangles. The anti-unitary elements of the complement $\hat{t}(p3m1-p3)$ are combinations of time reversal and mirror operation: a pure mirror operation parallel to $\mathbf{M}$ would reverse $\mathbf{M}$. These anti-mirror planes (and anti-glide planes) are represented by continuous and dashed red lines respectively. Operating the unitary and anti-unitary point group operators of the symmetry elements on an arbitrary $\mathbf{k}-$vector leads to the star representation as shown in Fig.\ref{figlayout}(b). The absence of time reversal combined with the symmetry reduction of the elementary motif limits the number of branches in the $\mathbf{k}$-star to 6 \textit{without} inversion center in k-space (instead of 12 for a symmetric hexagonal lattice). In consequence, the band structure is non-reciprocal and the irreducible Brillouin zone (IBZ) is larger than that of a symmetric hexagonal PhC. A possible choice for the IBZ is the shaded area in Fig.\ref{figlayout}(b), which now includes both a general Bloch mode and its time reversed counterpart. Indeed, because of the anti-unitary character of the mirror symmetry elements in the $p3m'1$ group, $-\mathbf{k}$ is equivalent to $m_{xy}\mathbf{k}$.
\begin{equation}
-\mathbf{k}= \hat{t}\mathbf{k} \thicksim \hat{t}m_{xy}'\mathbf{k} \thicksim \hat{t}\hat{t}m_{xy}\mathbf{k}\thicksim m_{xy}\mathbf{k}
\end{equation}
Except for the Bloch vectors $\mathbf{k}$ on the $\Gamma-M$ line (parallel to the anti-mirrors in $p3m'1$), all $\mathbf{k}$-points in the IBZ are intrinsically non-reciprocal. This explains the difference in notation between the $K$ and $K'$ points in the IBZ. The non-reciprocity and possible unidirectionality will thus be more pronounced the further one moves away from $\Gamma-M$. The highest effects are then expected around the $K$-point (and the non-equivalent time reversed $K'$ point). Using only general arguments from magnetic group theory we have with this specific example shown how a candidate MO PhC layout for one-way effects can be identified and where in the photonic bands the strongest non-reciprocal effects can be found. It suffices to choose a suitable magnetic group that doesn't possess any inversion symmetry in $\mathbf{k}$-space and then use a geometry for the elementary motif that is in accordance with the chosen magnetic group. As illustrated by the example, this can be achieved with just a slight geometric symmetry reduction.

\begin{figure*}[!tb]
\includegraphics[width=\textwidth]{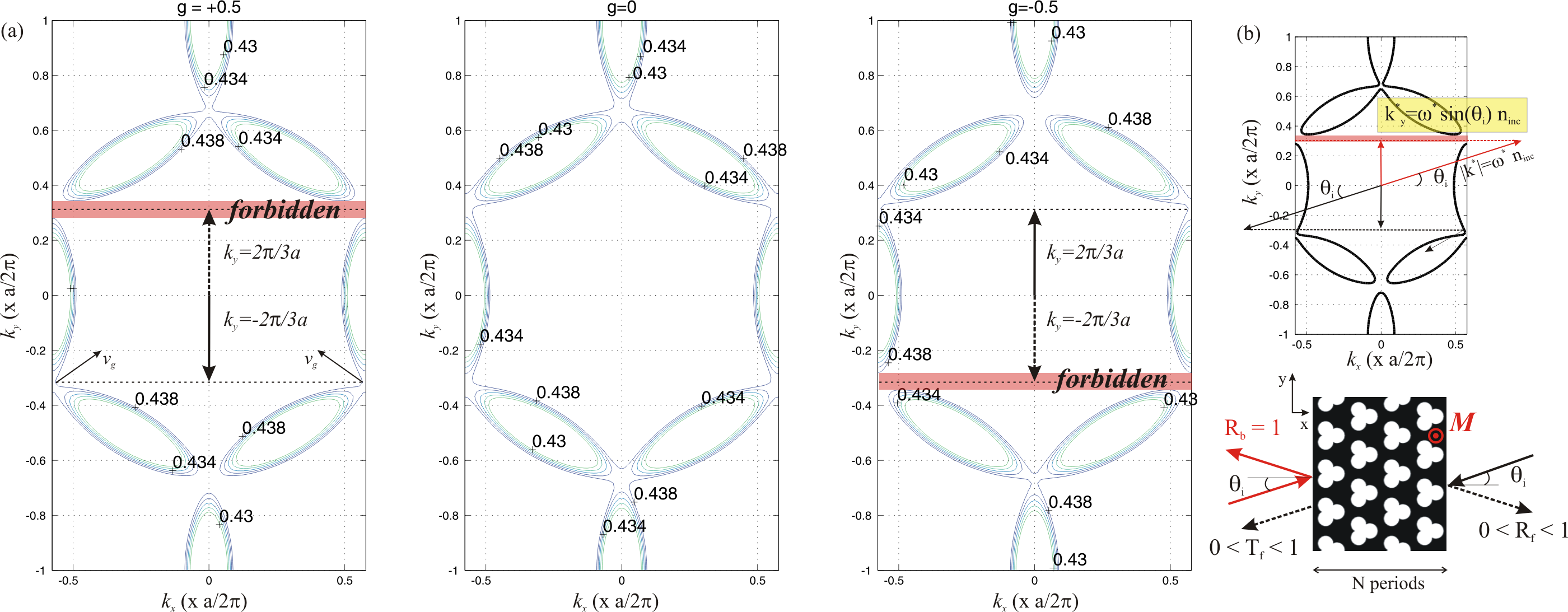}
\caption{(a) TE isofrequency contours for $0.43 \leqslant \omega^{*} \leqslant 0.438$ in reciprocal space for the MOPhC of Fig.\ref{figlayout}(a) with, from left to right, $g = 0.5, 0, -0.5$ . The strong non-reciprocity of the band structure around the $K(K')$ points induces a local forbidden zone for Bloch vectors with $k_y \approx \frac{2\pi}{3a}$ for $g>0$ or the opposite for $g<0$. For a nonmagnetic PhC ($g=0$) with reduced-symmetry motif, the band structure preserves the hexagonal symmetry of the lattice and is reciprocal. (b) Principle of a unidirectional uniformly magnetized MOPhC mirror based on the one way $k-$space band gap. A TE plane wave ($E_z=0$) with $k_y\approx\frac{2\pi}{3a}$ and $\omega_K\leqslant\omega^*\leqslant\omega_{K'}$ (second TE band) will be totally reflected in one sense and partially transmitted in the opposite sense. The condition for the optimum incidence angle depends both on the frequency and the index of the incidence medium via $\omega^* n_{inc} \sin(\theta_i) = 1/3$.}\label{isofreqfig}
\end{figure*}
Let us now illustrate this non-reciprocity from band diagram calculations for the 2D MOPhC of Fig.\ref{figlayout}. The permittivity tensor of the MO material, for instance Bismuth iron garnet (BIG, Bi$_3$Fe$_5$O$_{12}$) is written as
\begin{equation}\label{epsilonBIG}
\overline{\overline{\epsilon}} = \epsilon_0
\begin{pmatrix} 
\epsilon & +jg_z & -jg_y\\
-jg_z & \epsilon & +jg_x\\
+jg_y & -jg_x & \epsilon\end{pmatrix}
\end{equation}
where $\epsilon$ is the isotropic permittivity of the material, with  $g_x, g_y$ and$g_z$ the components of the gyrotropy vector $\mathbf{g}= g \mathbf{1}_\mathbf{M}$, with $\mathbf{1}_\mathbf{M}$ the unit vector along the magnetization direction, and $g$ is the gyrotropy constant. This tensor form is an approximation that leaves out the crystalline description of the material, and the non-reciprocal anisotropy only stems from the direction of the magnetization. For most MO materials, this is an acceptable description. The gyrotropy constant $g$ is related to the specific Faraday rotation through $\Theta_F = \frac{\pi g}{n \lambda}$. We take $\epsilon = 5.0625$ and $g = 0.1$ for BIG at near infrared frequencies \cite{Tepper:JCG03}. The antisymmetry of the off-diagonal elements, i.e. the source of the non-reciprocity, is imposed by Onsager's principle ($\overline{\overline{\epsilon}}_{kl}(\mathbf{M})=\overline{\overline{\epsilon}}_{lk}(-\mathbf{M})$), while power conservation for transparent materials ($\epsilon $ and $ g \in \mathbb{R}$) imposes the hermiticity of $\overline{\overline{\epsilon}}$. The sign of the gyrotropy vector components $g_x,g_y$ and $g_z$ changes under time reversal since they are proportional to the magnetization unit vector. Besides, any mirror plane parallel to $z$ will induce a change of sign of the $g_z$. For the 2D crystal of Fig.\ref{figlayout}, $g_x = g_y = 0$, and $g_z = g$ is a constant throughout the uniformly magnetized MO material. Within these conditions, the 2D MOPhC is said to be in Voigt configuration \cite{Zvezdin:IOP}. The specific form of the permittivity tensor leaves the TE ($E_z = 0$) and TM ($E_x=E_y=0$) Bloch modes of the 2D crystal uncoupled, while it leads to a non-reciprocal correction of the TE mode dispersion via the antisymmetric elements ($\epsilon_{xy} = -\epsilon_{yx}$).

Fig.~\ref{isofreqfig}(a) shows isofrequency contours in the reciprocal space of the crystal for normalized frequencies of the second TE band near the $K$ and $K'$ points:  $0.34\leqslant\omega^* (\equiv \omega a / 2\pi c)\leqslant0.348$. These contours have been obtained using a 128$\times$128-planewave expansion of the unit cell of the periodic permittivity tensor profile \cite{Johnson:OE01}. The geometrical parameters of the crystal are $r/a=0.2$ and $d/a=0.2$. The three graphs differ only by the sign and/or value of the gyrotropy coefficient~$g$. The middle graph shows the contours when the external magnetic field is switched off ($g=0$). As expected, the bands are reciprocal and the isofrequency contours exhibit the hexagonal symmetry of the Bravais lattice, notwithstanding the reduced triangular symmetry of the motifs. On the contrary, as soon as the magnetization is switched on (left and right graphs in Fig.~\ref{isofreqfig}(a)), the time-reversal symmetry is broken, and the reduced symmetry of the motif manifests itself clearly in the shape of the isofrequency contours.  Incidentally, the left and the right graph, obtained for exactly opposite values of $g$, transform into each other by a $k$-space inversion, which is equivalent to time reversal. This corroborates the validity of our numerical method. Note that to highlight the non-reciprocity of the bands we took $g=\pm 0.5$, which is about five times greater than a realistic material constant.
The most striking consequence of the time-reversal symmetry breaking is the relative frequency shift of the second TE band near the $K$ and $K'$ points. For $g>0$, the frequency at $K$ is higher than the one at $K'$; the situation is reversed for $g<0$. This leads to the formation of local forbidden zones in the reciprocal space: light propagation is forbidden at normalized frequencies close to $\omega^*=0.4385$ and wavevector projections on $\Gamma - K$ close to $k_y=\frac{2\pi}{3a}$ (if $g > 0$) and $-\frac{2\pi}{3a}$ (if $g<0$).
  
A one-way mirror operation can thus be obtained for an incident TE plane wave at these frequencies provided that the interface between the photonic crystal and the incidence medium is aligned in the $\Gamma-K$ direction and that the $k_y$ component of the incident plane wave falls within the forbidden zone. The same plane wave incident from the opposite side under the same angle (i.e. with opposite $k_y$) will have finite transmission. This principle is schematized in Fig.\ref{isofreqfig}(b). The angular acceptance of the mirror is directly related to the width $\Delta k_y$ of the unidirectional gap in $k-$space. This width is itself dependent on the operation frequency and reaches its maximum at the nonreciprocal midgap frequency (halfway between the frequencies at the $K$ resp. $K'$ points). The useful bandwidth of the device can be identified with this maximum width. For realistic material parameters ($g=0.1$), one obtains $\Delta k^*_y\equiv\Delta k_y a/2 \pi \approx 10^{-3}$ at the midgap frequency. The normalized frequency bandwidth $\Delta\omega^*$ is of the order of $10^{-4}$. In a pessimistic approximation, the MO PhC slice thickness required for good non-reciprocal extinction is estimated to be: $L \sim 1/\Delta k_y \approx 150 a$, which represents about $40 \lambda$ at MO gap frequencies. The device bandwidth for a non-reciprocal extinction ratio larger than 100 is estimated to be about 100GHz at near infrared frequencies.
\begin{figure*}[!tb]
\begin{tabular}{c c c}
\includegraphics[width=0.33\textwidth]{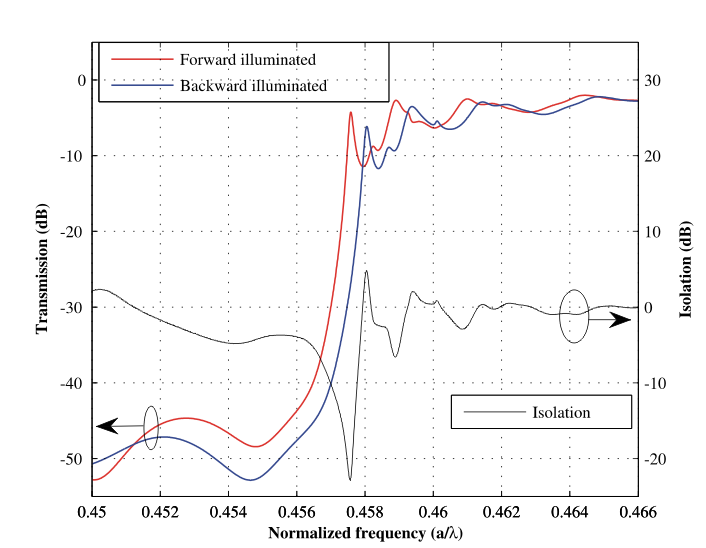}&\includegraphics[width=0.33\textwidth]{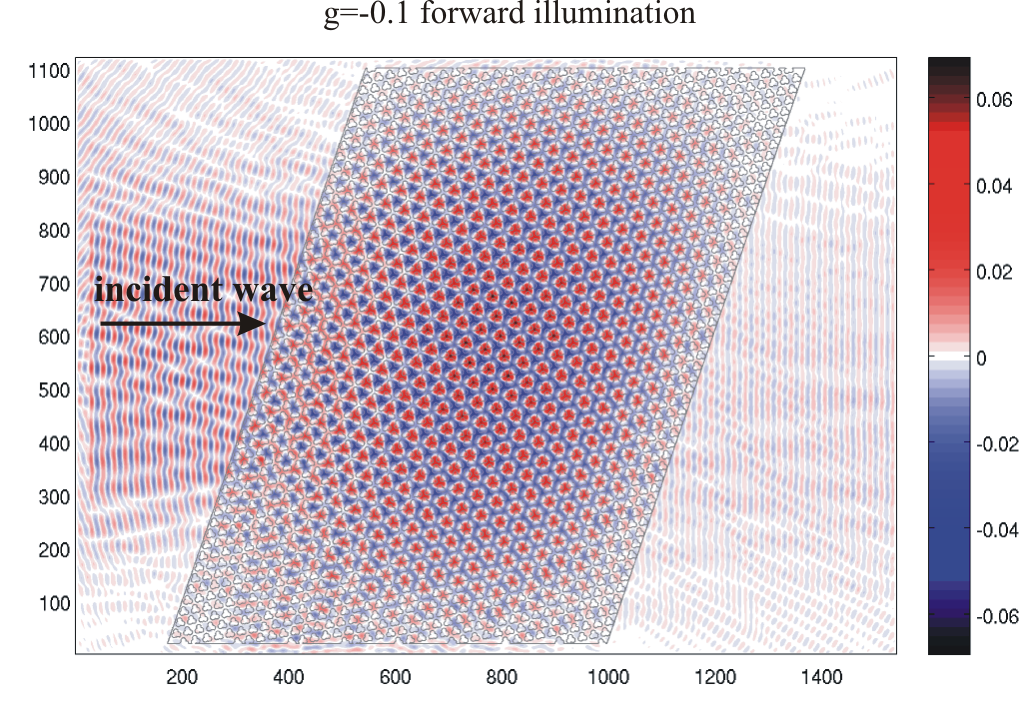}&\includegraphics[width=0.33\textwidth]{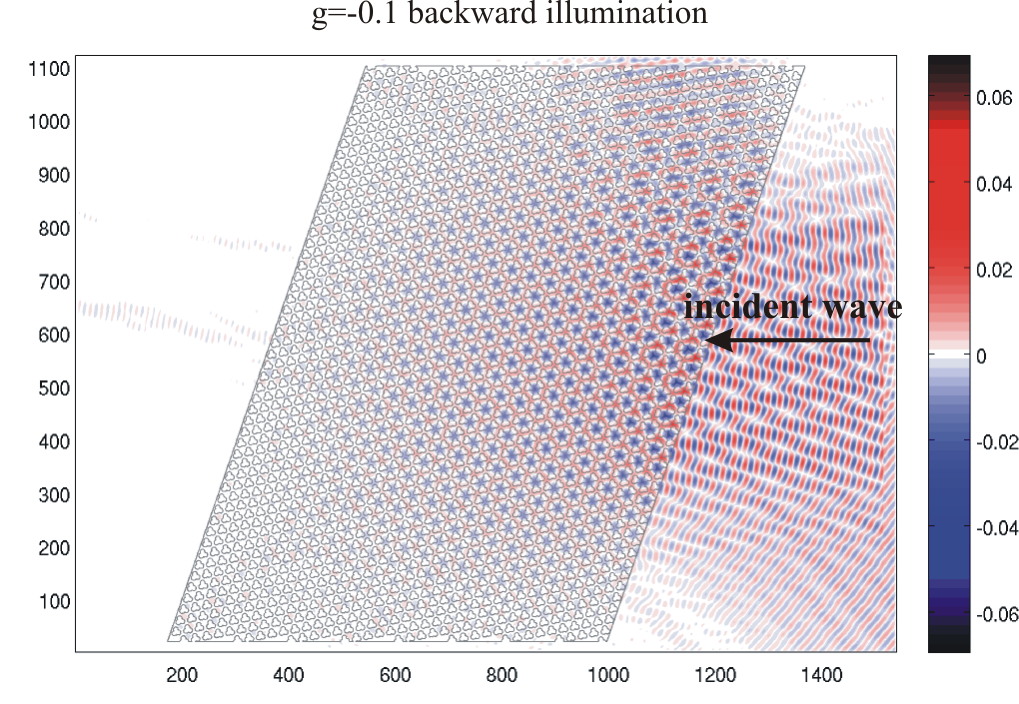}
\end{tabular}
\caption{(a): Normalized power transmission spectrum of the 40 period thick uniformly magnetized unidirectional MOPhC mirror calculated by FDTD in the frequency range of interest. The gray curve shows the non-reciprocal extinction ratio. A relative bandwidth $\frac{\Delta\omega^*}{\omega^*} \approx 10^{-3} $ is obtained for a one-way extinction ratio larger than 40. (b): $H_z$ steady-state field maps under TE plane wave illumination in forward and backward direction at $\omega^*=0.457$. In the forward direction light is resonantly transmitted while in the backward direction the unidirectional band gap rejects the incident light.\label{FDTD2}}
\end{figure*}
The unidirectional transmission of the mirror was verified using an in-house developed 2D FDTD (Finite Difference Time Domain Method) model with a tensorial description of the dielectric permittivity\cite{Checoury:APL07}. The results of the FDTD calculations are shown in Fig.\ref{FDTD2}. The mirror consists of a finite 40-period thick MOPhC slab embedded in a homogeneous isotropic medium with $\epsilon$ equal to that of the BIG background of the MOPhC. Both the PhC and MO material parameters are the same as in Fig. \ref{isofreqfig} except for $g$ whose value ($-0.1$) is more realistic. The mirror is tilted at an angle of $20^\circ$ (in accordance with the condition of Fig.\ref{isofreqfig}(b)), and is illuminated by a TE plane wave. Fig.\ref{FDTD2}(a) shows the normalized power transmission spectrum under forward and backward illumination and with a uniform $-z$ magnetization ($g=-0.1$). This corresponds to the situation in the right graph of Fig.\ref{isofreqfig}(a). The non-reciprocal transmission is clearly observed. A one-way transmission with a non-reciprocal power extinction ratio close to 400 is obtained in the presence of uniform magnetization. It may be noticed that the normalized midgap frequency is about 2\% higher than that calculated from the plane wave expansion method (Fig.\ref{isofreqfig}). This discrepancy is believed to be due to different resolutions used in the two methods. A visual illustration of the unidirectional light rejection is given in Fig.\ref{FDTD2}(b-c), which shows the $H_z$ field distribution of a continuous ($\omega^*=0.457$) TE plane wave incident either in the forward or the backward direction. The light rejection in the backward direction is striking.

In conclusion, we have demonstrated a general approach to obtain 2D MO PhC structures that present optical one-way gaps. Using concepts of magnetic group theory we have thus proven that strong non-reciprocal on-off properties are not necessarily limited to structures with antiparallel magnetizations in the unit cell. Instead of that, the use of a motif with slightly reduced symmetry suffices to achieve strong nonreciprocal light rejection in the presence of uniform magnetization. The use of antiparallel magnetic domains is just another specific case among the possible realizations of a non-reciprocal magnetic group. Shubnikov has shown that in two dimensions magnetic periodic structures can be classified into 46 distinct groups \cite{Shubnikovbook}. Among these only five groups do not allow an inversion of $\mathbf{k}$-space with uniform magnetization. One of these five groups is the $p3m'1$ investigated in this work. From a fundamental point of view this work is thus believed to open the way to a generic scheme for one-way optics in 2D photonic crystals.

This work was supported by the French Research Agency through the MAGNETOPHOT project.


\end{document}